\documentclass[11pt]{article}
\usepackage{amssymb,latexsym,amsmath}
\usepackage[dvips]{graphicx}
\usepackage{cite}
\newtheorem{theo}{Theorem}
\newtheorem{defi}[theo]{Definition}

\newtheorem{prop}[theo]{Proposition}
\def\nn{\nonumber}
\def\Z{{\mathbb Z}}

\def\R{{\mathbb R}}

\def\su{\mathfrak{su}}
\def\u{\mathfrak{u}}
\def\so{\mathfrak{so}}
\def\osp{\mathfrak{osp}}
\newcommand{\myatop}[2]{\genfrac{}{}{0pt}{}{#1}{#2}}

\begin{document}
\begin{center}
{\Large \bf
Finite oscillator models: the Hahn oscillator}\\[5mm]
{\bf E.I.\ Jafarov\footnote{Permanent address: 
Institute of Physics, Azerbaijan National Academy of Sciences, Javid av.\ 33, AZ-1143 Baku, Azerbaijan}, 
N.I.~Stoilova\footnote{Permanent address: 
Institute for Nuclear Research and Nuclear Energy, Boul.\ Tsarigradsko Chaussee 72,
1784 Sofia, Bulgaria} 
and J.\ Van der Jeugt} \\[1mm]
Department of Applied Mathematics and Computer Science,
Ghent University,\\
Krijgslaan 281-S9, B-9000 Gent, Belgium\\[1mm]
E-mail: ejafarov@physics.ab.az, Neli.Stoilova@UGent.be and 
Joris.VanderJeugt@UGent.be
\end{center}

\vskip 10mm
\noindent
Short title: Hahn oscillator



\begin{abstract}
A new model for the finite one-dimensional harmonic oscillator is proposed based upon the algebra $\u(2)_\alpha$.
This algebra is a deformation of the Lie algebra $\u(2)$ extended by a parity operator, with deformation parameter $\alpha$.
A class of irreducible unitary representations of $\u(2)_\alpha$ is constructed.
In the finite oscillator model, the (discrete) spectrum of the position operator is determined,
and the position wavefunctions are shown to be dual Hahn polynomials.
Plots of these discrete wavefunctions display interesting properties, similar to those of the parabose oscillator.
We show indeed that in the limit, when the dimension of the representations goes to infinity,
the discrete wavefunctions tend to the continuous wavefunctions of the parabose oscillator. 
\end{abstract}

\section{Introduction}

Finite oscillator models obey the same dynamics as the classical and quantum oscillators, but the 
operators corresponding to position, momentum and Hamiltonian are elements of some algebra different
from the traditional oscillator Lie algebra.
The interest in finite oscillator models comes primarily from optical image processing~\cite{Atak2005}.
In this context, most attention has been paid to a finite oscillator model based on the Lie algebra
$\su(2)$ (or $\so(3)$), generalized to $\so(4)$ in the case of a finite two-dimensional oscillator~\cite{Atak2001,Atak2001b,Atak2005}.

In the one-dimensional case, the setting is as follows.
There are three (essentially self-adjoint) operators: a position operator $\hat q$, its corresponding momentum operator $\hat p$ and
a (pseudo-) Hamiltonian $\hat H$ which is the generator of time evolution. 
These operators should satisfy the Hamiltonian-Lie equations (or the compatibility of Hamilton's equations with the Heisenberg
equations):
\begin{equation}
[\hat H, \hat q] = -i \hat p, \qquad [\hat H,\hat p] = i \hat q,
\label{Hqp}
\end{equation}
in units with mass and frequency both equal to~1, and $\hbar=1$.
Contrary to the canonical case, the commutator $[\hat q, \hat p]=i$ is not required. 
Wigner considered such a system already in 1950~\cite{Wigner}.
He required the extra condition
\begin{equation}
\hat H = \frac12(\hat p^2 + \hat q^2).
\label{H-W}
\end{equation}
In that case, one is dealing with the Wigner quantum oscillator (or parabose oscillator)~\cite{Palev79,Palev82}.
In fact, the algebraic structure equivalent with~\eqref{Hqp} and~\eqref{H-W} is just the Lie superalgebra $\osp(1|2)$ (see Appendix).
It turns out that in this case one is still dealing with an oscillator with an infinite energy spectrum, the only difference being
that this spectrum is shifted compared to the canonical case. This shift is determined by the $\osp(1|2)$ representation
parameter $a$ ($a$ is real and positive), the spectrum being $n+a$ ($n=0,1,2,\ldots$).

If one wants a finite oscillator model, the relation~\eqref{H-W} should be dropped and one is left with~\eqref{Hqp} only.
On top of~\eqref{Hqp} and the self-adjointness, it is then common to require the following conditions~\cite{Atak2001}:
\begin{itemize}
\item all operators $\hat q$, $\hat p$, $\hat H$ belong to some (Lie) algebra (or superalgebra) $\cal A$;
\item the spectrum of $\hat H$ in (unitary) representations of $\cal A$ is equidistant.
\end{itemize}
The first requirement is often taken to be stronger, in the sense that all commutator brackets between operators close into a Lie
algebra (or its universal enveloping algebra)~\cite{Atak2001}; it is also often taken to be weaker, in the sense that the
remaining commutator $[\hat q, \hat p]$ is just required to be a ``function'' of $\hat H$~\cite{Arik1999}.
Also the second requirement is sometimes dropped~\cite{Arik1999}.

The case with ${\cal A}= \su(2)$ (or its enveloping algebra) has been treated extensively in a number of papers~\cite{Atak2001,Atak2001b,Atak2005}.
In that case, the relevant representations are the common $\su(2)$ representations labelled by an integer of half-integer $j$.
Up to a constant, the Hamiltonian $\hat H$ is the diagonal $\su(2)$ operator with a linear spectrum $n+\frac12$ ($n=0,1,\ldots,2j$).
Obviously also $\hat q$ and $\hat p$ have a finite spectrum. In this model it is given by $\{-j,-j+1,\ldots,+j\}$~\cite{Atak2001}.
More interestingly, the position wavefunctions have been constructed, and are given by Krawtchouk functions (normalized
symmetric Krawtchouk polynomials). 
These discrete wavefunctions have interesting properties, and their shape is reminiscent of those of the canonical oscillator~\cite{Atak2001}.
It was indeed shown that under the limit $j\rightarrow \infty$ the discrete wavefunctions coincide with the 
canonical wavefunctions in terms of Hermite polynomials~\cite{Atak2001,Atak2003}.

In the present paper, we construct a one-parameter deformation of the enveloping algebra of $\u(2)$, with a real deformation
parameter $\alpha>-1$. This algebra, denoted by $\u(2)_\alpha$, is chosen as the algebra $\cal A$ to which the operators
$\hat q$, $\hat p$ and $\hat H$ should belong, along with the relations~\eqref{Hqp}. 
For the value $\alpha=-\frac12$, $\u(2)_\alpha$ reduces to the undeformed $\u(2)$ (or $\su(2)$).
For half-integer $j$ values (not for integer $j$-values), the common $\su(2)$ or $\u(2)$ representations can be
extended to representations of $\u(2)_\alpha$.
An interesting aspect of these representations, is that the operators $\hat q$ and $\hat p$ have again a simple
spectrum. 
Furthermore, the position wavefunctions can be computed explicitly.
These wavefunctions turn out to be dual Hahn functions (normalized dual Hahn polynomials). 
The dual Hahn polynomials, usually characterized by two parameters $(\alpha,\beta)$, appear in the wavefunctions
with parameters $(\alpha,\alpha+1)$ or $(\alpha+1,\alpha)$, depending on the parity of the wavefunction.
We present some plots of these discrete wavefunctions, which have interesting properties similar to those
of the $\su(2)$ finite oscillator model (to which they reduce when $\alpha=-\frac12$).
To understand these properties, the limit $j\rightarrow\infty$ is determined.
Quite surprisingly, under this limit the discrete wavefunctions become the parabose oscillator wavefunctions.
So, the finite oscillator model presented here could be interpreted as a finite parabose oscillator model,
which reduces to the $\su(2)$ finite oscillator model when $\alpha=-\frac12$.

The structure of the paper is as follows: in Section~2 we construct the algebra $\u(2)_\alpha$ and 
its representations. In Section~3, it is shown that $\u(2)_\alpha$ gives rise to a new finite oscillator model.
The main result here is the determination of the spectrum of the position operator $\hat q$, and the
computation of its eigenvectors.
In Section~4, we present the position wavefunctions of the $\u(2)_\alpha$ oscillator model.
We discuss some of their properties, and give the limit relation to parabose wavefunctions.
Finally, we have included an Appendix where some aspects of the parabose oscillator (or Wigner quantum oscillator)
are summarized.

\section{The algebra $\u(2)_\alpha$ and its representations}

The Lie algebra $\su(2)$~\cite{Wybourne,Humphreys} is usually defined by its basis elements 
$J_0$, $J_+$, $J_-$ with commutators $[J_0,J_\pm]=\pm J_\pm$ and $[J_+,J_-]=2J_0$.
The non-trivial unitary representations of $\su(2)$, corresponding to the star relations $J_0^\dagger=J_0$, 
$J_\pm^\dagger=J_\mp$, are labelled~\cite{Wybourne,Humphreys} by a positive integer or half-integer $j$. These representations have dimension $2j+1$,
and the action of the $\su(2)$ operators on a set of basis vectors $|j,m\rangle$ (with $m=-j,-j+1,\ldots,+j$) is given by
\[
J_0 |j,m\rangle = m\;|j,m\rangle,\qquad 
J_\pm |j,m\rangle = \sqrt{(j\mp m)(j\pm m +1)}\;|j,m\pm 1\rangle.
\]
The Lie algebra $\su(2)$ can be extended to $\u(2)$ by adding an extra operator $C$ which commutes with all basis elements $J_0$, $J_+$, $J_-$;
the action of $C$ in the above representations is diagonal: $C |j,m\rangle = (2j+1)\;|j,m\rangle$.
It is also possible to extend $\u(2)$ further by a parity operator $P$, whose action in these representations is given by
$P |j,m\rangle = (-1)^{j+m}\;|j,m\rangle$. 
On the algebraic level, this means that we extend the universal enveloping algebra of $\u(2)$ by an operator $P$ 
that commutes with $C$ and $J_0$, that anticommutes with $J_+$ and $J_-$, and for which $P^2=1$.

The addition of these two operators does not yet give much extra structure as far as representations are concerned.
But this extended algebra can be deformed by a parameter $\alpha$, leading us to the definition of $\u(2)_\alpha$.

\begin{defi}
Let $\alpha$ be a parameter. The algebra $\u(2)_\alpha$ is a unital algebra with basis elements $J_0$, $J_+$, $J_-$, $C$ and $P$ subject to 
the following relations:
\begin{itemize}
\item $C$ commutes with all basis elements,
\item $P$ is a parity operator satisfying $P^2=1$ and
\begin{equation}
[P,J_0]=PJ_0-J_0P=0, \qquad \{P,J_\pm\}=PJ_\pm + J_\pm P= 0.
\label{P}
\end{equation}
\item The $\su(2)$ relations are deformed as follows:
\begin{align}
& [J_0, J_\pm] = \pm J_\pm,  \label{J0J+} \\
& [J_+, J_-] = 2 J_0 - (2\alpha+1)^2P - (2\alpha+1) CP.
\label{J+J-}
\end{align}
\end{itemize}
\end{defi}
Relation~\eqref{J+J-} looks at this moment fairly arbitrary. 
It would also be of little interest, if the representations of $\u(2)$ could not be extended to representations of $\u(2)_\alpha$;
for the current choice of~\eqref{J+J-}, this happens to be the case.
Note that for $\alpha=-\frac12$, \eqref{J+J-} reduces to the usual $\su(2)$ commutator relation; so for $\alpha=-\frac12$ the deformation is trivial.

The extension of the common $(2j+1)$-dimensional representations of $\u(2)$ to representations of $\u(2)_\alpha$ is
possible provided $j$ is a half-integer (and not when $j$ is an integer). 
This is presented in the following proposition.

\begin{prop}
Let $j$ be a half-integer (i.e.\ $2j$ is odd), and consider the space $V_j$ with basis vectors 
$|j,-j\rangle$, $|j,-j+1\rangle$, $\ldots$, $|j,j\rangle$. Assume that $\alpha>-1$. Then the following action turns $V_j$ into
an irreducible representation space of $\u(2)_\alpha$.
\begin{align}
& C |j,m\rangle = (2j+1)\;|j,m\rangle, \label{act-C}\\
& P |j,m\rangle = (-1)^{j+m}\;|j,m\rangle,\label{act-P}\\
& J_0 |j,m\rangle = m\;|j,m\rangle,\label{act-J0}\\
& J_+ |j,m\rangle = 
  \begin{cases}
 \sqrt{(j-m)(j+m +1)}\;|j,m+1\rangle, & \text{if $j+m$ is odd;}\\
 \sqrt{(j-m+2\alpha+1)(j+m+2\alpha+2)}\;|j,m+1\rangle, & \text{if $j+m$ is even.}
 \end{cases} \label{act-J+}\\
& J_- |j,m\rangle = 
  \begin{cases}
 \sqrt{(j+m)(j-m +1)}\;|j,m-1\rangle, & \text{if $j+m$ is even;}\\
 \sqrt{(j+m+2\alpha+1)(j-m+2\alpha+2)}\;|j,m-1\rangle, & \text{if $j+m$ is odd.}
 \end{cases} \label{act-J-}
\end{align} 
\end{prop}

Note that, for a general parameter $\alpha$, the condition $2j$ odd is necessary in order to have $J_+|j,j\rangle=0$ and $J_-|j,-j\rangle=0$.
The condition $\alpha>-1$ ensures that the factors under the square roots are positive ($\alpha=-1$ would lead to reducibility).
To actually prove the proposition it is sufficient to verify that all defining relations from Definition~1 are 
valid when acting on an arbitrary vector $|j,m\rangle$, and this is a straightforward calculation.
Irreducibility follows e.g.\ from the fact that $(J_+)^k |j,-j\rangle$ is nonzero and proportional to $|j,-j+k\rangle$ for $k=1,2,\ldots,2j$,
and similarly $(J_-)^k |j,j\rangle$ is nonzero.

Note also that the representation given in this proposition is unitary under the star conditions $C^\dagger =C$, $P^\dagger=P$, $J_0^\dagger=J_0$, 
$J_\pm^\dagger=J_\mp$.

\section{A one-dimensional oscillator model based on $\u(2)_\alpha$}

The conditions for finite oscillator models have been described in Section~1.
In the context of the algebra $\u(2)_\alpha$, one can make the following choice for the position, momentum and Hamiltonian operator:
\begin{equation}
\hat q = \frac12 (J_++J_-), \qquad
\hat p = \frac{i}{2}(J_+-J_-), \qquad
\hat H = J_0+\frac12 C.
\end{equation}
It is easy to verify that~\eqref{Hqp} is satisfied.
Furthermore, in the representation space $V_j$, $\hat H |j,m\rangle = (m+j+\frac12)|j,m\rangle$, so the spectrum of $\hat H$ is 
indeed linear and given by
\begin{equation}
n+\frac12 \qquad (n=0,1,\ldots,2j).
\end{equation}

An interesting question is the determination of the eigenvalues of the operators $\hat q$ (position eigenvalues)
and $\hat p$ (momentum eigenvalues), and the corresponding eigenvectors.
We shall perform this for the position operator (for $\hat p$, the analysis is similar).

{}From the actions~\eqref{act-J+}-\eqref{act-J-}, one finds
\begin{equation*}
2\hat q |j,m\rangle = \sqrt{(j+m)(j-m +1)}\;|j,m-1\rangle +
\sqrt{(j-m+2\alpha+1)(j+m+2\alpha+2)}\;|j,m+1\rangle
\end{equation*}
if $j+m$ is even, and
\begin{equation*}
2\hat q |j,m\rangle = \sqrt{(j+m+2\alpha+1)(j-m+2\alpha+2)}\;|j,m-1\rangle +
\sqrt{(j-m)(j+m +1)}\;|j,m+1\rangle
\end{equation*}
if $j+m$ is odd.
This means that in the (ordered) basis ${\cal B}$ of $V_j$,
\begin{equation}
{\cal B} = \{ |j,-j\rangle, |j,-j+1\rangle, \ldots, |j,j-1\rangle, |j,j\rangle \},
\end{equation}
the operator $2\hat q$ takes the matrix form
\begin{equation}
2\hat q=\left(
\begin{array}{ccccc}
0 & M_0& 0 & \cdots & 0 \\
M_0 & 0 & M_1 & \cdots & 0\\
0 & M_1 & 0 & \ddots &  \\
\vdots & \vdots & \ddots & \ddots& M_{2j-1}\\
0 & 0 &  & M_{2j-1} & 0
\end{array}
\right),
\label{M}
\end{equation}
with
\begin{equation}
M_k= \begin{cases}
 \sqrt{(k+1)(2j-k)}, & \text{if $k$ is odd;}\\
 \sqrt{(k+2\alpha+2)(2j-k+2\alpha+1)}, & \text{if $k$ is even.}
 \end{cases}
\label{Ma}
\end{equation}
For this matrix, the eigenvalues are known explicitly~\cite{Shi2005,SV2011}.
\begin{prop}
The $2j+1$ eigenvalues $q$ of the position operator $\hat q$ in the representation $V_j$ are given by
\begin{equation}
-\alpha-j-\frac12, -\alpha-j+\frac12, \ldots, -\alpha-1; \alpha+1,\alpha+2, \ldots,\alpha+j+\frac12.
\end{equation}
It will be appropriate to label these $\hat q$-eigenvalues as $q_k$, where $k=-j,-j+1,\ldots,+j$, so
\[
q_{\pm k} = \pm(\alpha+k+\frac12), \qquad k=\frac12, \frac32, \ldots, j.
\]
\end{prop}
In the $\su(2)$ oscillator model~\cite{Atak2001}, the spectrum of $\hat q$ in the representation $V_j$ with $2j$ odd is given by
\[
-j, -j+1, \ldots, -\frac12; \frac12, \ldots,j+1, j.
\]
So in the $\u(2)_\alpha$ oscillator model, there is a shift from the origin by $\alpha+\frac12$ (so for positive 
$\hat q$-eigenvalues a shift by $\alpha+\frac12$; and for negative $\hat q$-eigenvalues a shift by $-(\alpha+\frac12)$).
Note that for $\alpha>-\frac12$ this shift is away from the origin; for $-1<\alpha<-\frac12$ this shift is towards the origin.

In a recent paper~\cite{SV2011}, it was shown that also the eigenvectors of a tridiagonal matrix of the form~\eqref{M} can be
constructed explicitly. 
The expressions of these eigenvectors involve Hahn polynomials, so let us first recall some notation.
Hahn polynomials $Q_n(x;\alpha, \beta, N)$~\cite{Koekoek,Suslov} of degree $n$ ($n=0,1,\ldots,N$) in the variable $x$, with parameters 
$\alpha>-1$ and $\beta>-1$ are defined by~\cite{Koekoek,Suslov}:
\begin{equation}
Q_n(x;\alpha,\beta,N) = {\;}_3F_2 \left( \myatop{-n,n+\alpha+\beta+1,-x}{\alpha+1,-N} ; 1 \right),
\label{defQ}
\end{equation}
in terms of the generalized hypergeometric series $_3F_2$ of unit argument~\cite{Bailey,Slater}. 
Hahn polynomials satisfy a (discrete) orthogonality relation~\cite{Koekoek}:
\begin{equation}
\sum_{x=0}^N w(x;\alpha, \beta,N) Q_l(x;\alpha, \beta, N) Q_n(x;\alpha,\beta,N) = h(n;\alpha,\beta,N)\, \delta_{ln},
\label{orth-Q}
\end{equation} 
where
\begin{align*}
& w(x;\alpha, \beta,N) = \binom{\alpha+x}{x} \binom{N+\beta-x}{N-x} \qquad (x=0,1,\ldots,N); \\
& h(n;\alpha,\beta,N)= \frac{(n+\alpha+\beta+1)_{N+1}(\beta+1)_n n!}{(2n+\alpha+\beta+1)(\alpha+1)_n(N-n+1)_n N!}.
\end{align*}
We have used here the common notation for Pochhammer symbols~\cite{Bailey,Slater}
$(a)_k=a(a+1)\cdots(a+k-1)$ for $k=1,2,\ldots$ and $(a)_0=1$.
As $w$ is the weight function and $h(n;\alpha,\beta,N)$ the ``squared norm'', orthonormal Hahn functions $\tilde Q$ are determined by:
\begin{equation}
\tilde Q_n(x;\alpha,\beta,N) \equiv \frac{\sqrt{w(x;\alpha,\beta,N)}\, Q_n(x;\alpha,\beta,N)}{\sqrt{h(n;\alpha,\beta,N)}}.
\label{Q-tilde}
\end{equation}
Dual Hahn polynomials are almost the same as Hahn polynomials, but with the role of $x$ and $n$ in~\eqref{defQ} interchanged.
So for $x\in\{0,1,\ldots,N\}$, the right hand side of~\eqref{defQ} can be seen as the dual Hahn polynomial of degree $x$
in the variable $\lambda(n)=n(n+\alpha+\beta+1)$~\cite{Koekoek}. 

The next result follows now from~\cite[Proposition~2]{SV2011}:
\begin{prop}
The orthonormal eigenvector of the position operator $\hat q$ in $V_j$ for the eigenvalue $q_k$, denoted by $|j,q_k)$, is given 
in terms of the basis ${\cal B}$ by
\begin{equation}
|j,q_k) = \sum_{m=-j}^j U_{j+m,j+k} |j,m\rangle.
\end{equation}
Herein, $U=(U_{rs})_{0\leq r,s \leq 2j}$ is a $(2j+1)\times(2j+1)$ matrix with elements
\begin{align}
& U_{2r,j-s-\frac12} = U_{2r,j+s+\frac12} = \frac{(-1)^r}{\sqrt{2}} \tilde Q_s(r;\alpha,\alpha+1,j-\frac12), \label{Ueven}\\
& U_{2r+1,j-s-\frac12} = -U_{2r+1,j+s+\frac12} = -\frac{(-1)^r}{\sqrt{2}} \tilde Q_s(r;\alpha+1,\alpha,j-\frac12), \label{Uodd}
\end{align}
where $r,s\in\{0,1,\ldots,j-\frac{1}{2}\}$.
The functions $\tilde Q$ are normalized Hahn polynomials~\eqref{Q-tilde}.
\end{prop}
Note that $U$ is an orthogonal matrix, $UU^T=U^TU=I$, hence the $\hat q$ eigenvectors are orthonormal:
\[
(j, q_k | j, q_l) = \delta_{kl}.
\]

\section{$\u(2)_\alpha$ oscillator wavefunctions and their properties}

In general, the wavefunctions are the overlaps between the normalized eigenstates of the position operator and
the eigenstates of the Hamiltonian.
So the wavefunctions of the $\u(2)_\alpha$ finite oscillator are the overlaps between the $\hat q$-eigenvectors
and the $\hat H$-eigenvectors (or equivalently, the $J_0$-eigenvectors $|j,m\rangle$).
They are denoted by $\Phi^{(\alpha)}_{j+m}(q)$, where $m=-j,-j+1,\ldots,+j$, and where $q$ assumes one of the 
discrete values $q_k$ $(k=-j,-j+1,\ldots,+j)$. 
Concretely, following the notation of the previous section:
\begin{equation}
\Phi^{(\alpha)}_{j+m}(q_k)= \langle j,m | j,q_k ) = U_{j+m,j+k}.
\end{equation}

Let us examine this function in more detail. Since $U$ has a different form for even and odd indices, \eqref{Ueven} and \eqref{Uodd},
we shall also make this distinction here.
For $j+m$ even, $j+m=2n$, this is by~\eqref{Ueven} an even function of the position variable $q$; for positive $q$-values one has
\begin{equation*}
\Phi^{(\alpha)}_{2n} (q_{k}) = \frac{(-1)^n}{\sqrt{2}} \tilde Q_{k-\frac12}(n;\alpha,\alpha+1,j-\frac12), \qquad k=\frac12, \frac32, \ldots, j;
\end{equation*}
or more explicitly:
\begin{equation}
\Phi^{(\alpha)}_{2n} (q_{k}) = \frac{(-1)^n}{\sqrt{2}} 
\sqrt{\frac{w(n;\alpha,\alpha+1,j-\frac12)}{h(q_k-\alpha-1;\alpha,\alpha+1,j-\frac12)}} 
{\ }_3F_2 \left( \myatop{-q_k+\alpha+1,q_k+\alpha+1,-n}{\alpha+1,-j+\frac12} ; 1 \right).
\label{Phi-even}
\end{equation}
So one can interpret this as a dual Hahn polynomial of degree $n$ in the position variable $q_k$ 
(or rather in the variable $\lambda(k-\frac12)$).
For $j+m$ odd, $j+m=2n+1$, this is by~\eqref{Uodd} an odd function of the variable $q$; for positive $q$-values we have
\begin{equation*}
\Phi^{(\alpha)}_{2n+1} (q_{k}) = \frac{(-1)^n}{\sqrt{2}} \tilde Q_{k-\frac12}(n;\alpha+1,\alpha,j-\frac12) ,\qquad k=\frac12, \frac32, \ldots, j;
\end{equation*}
or explicitly:
\begin{equation}
\Phi^{(\alpha)}_{2n+1} (q_{k}) = \frac{(-1)^n}{\sqrt{2}} 
\sqrt{\frac{w(n;\alpha+1,\alpha,j-\frac12)}{h(q_k-\alpha-1;\alpha+1,\alpha,j-\frac12)}} 
{\ }_3F_2 \left( \myatop{-q_k+\alpha+1,q_k+\alpha+1,-n}{\alpha+2,-j+\frac12} ; 1 \right).
\label{Phi-odd}
\end{equation}
again a dual Hahn polynomial of degree $n$ in the position variable.
Because of the appearance of these polynomials as wavefunctions of the $\u(2)_\alpha$ finite oscillator,
we shall refer to this model as the Hahn oscillator.

\begin{figure}[htb]
\begin{tabular}{cc}
\hline\\[-3mm]
\includegraphics[scale=0.6]{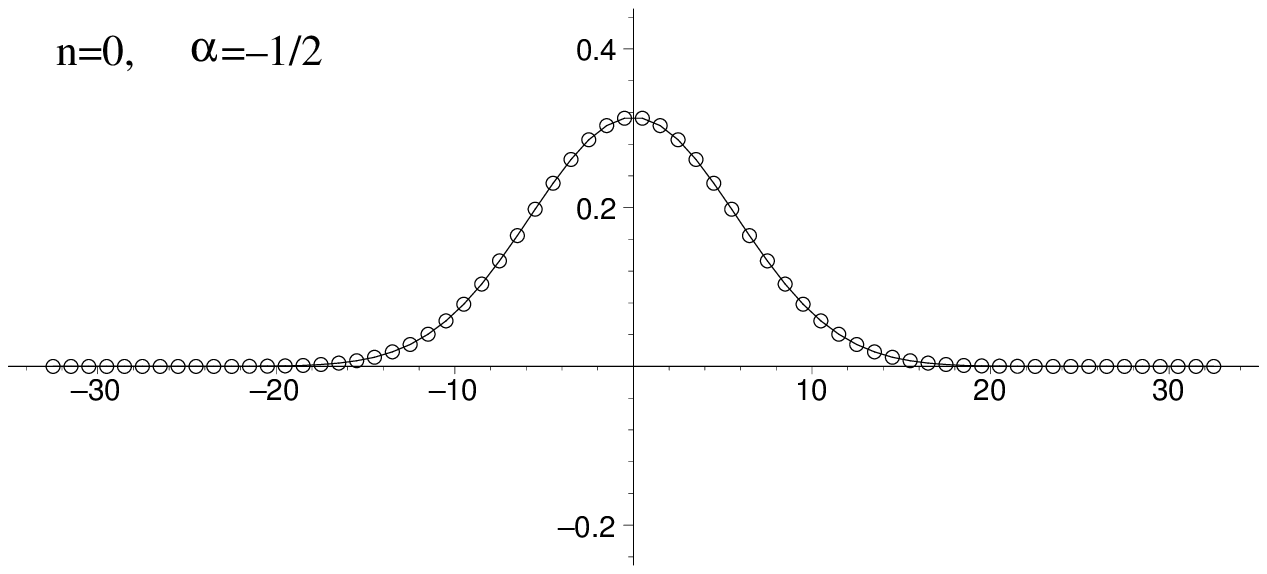} & \includegraphics[scale=0.6]{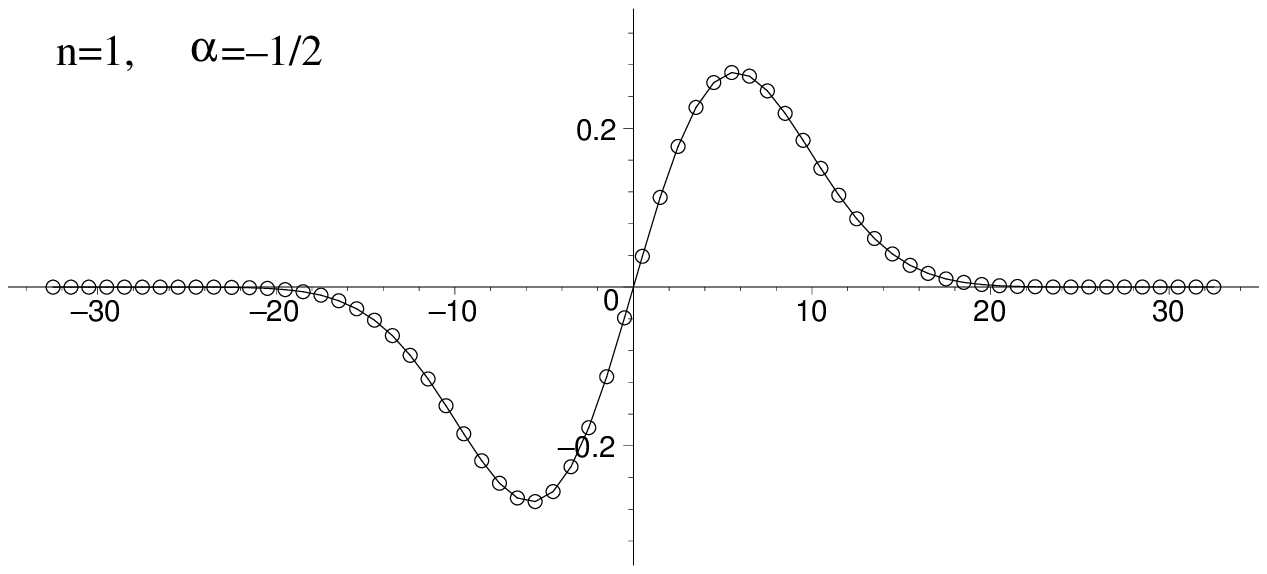} \\
\includegraphics[scale=0.6]{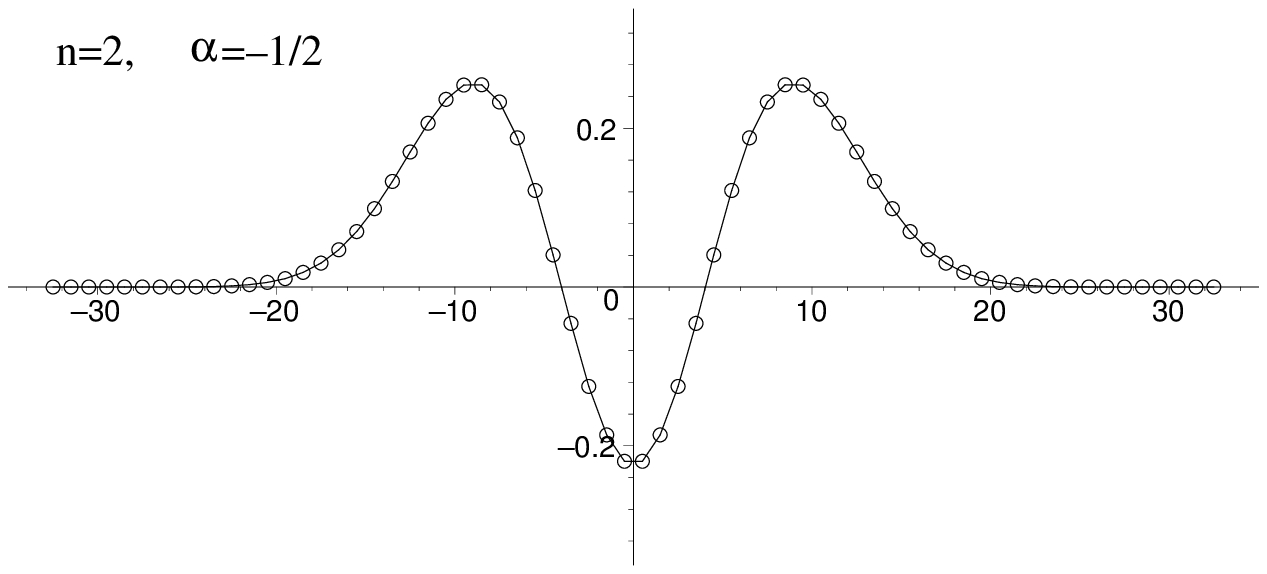} & \includegraphics[scale=0.6]{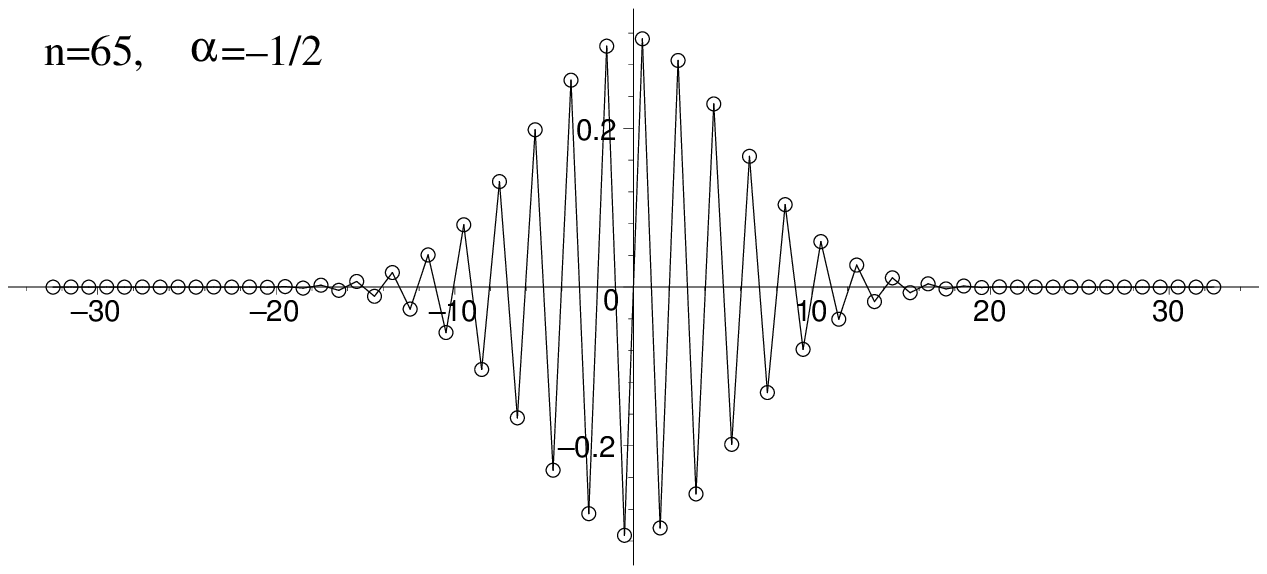} \\[-1mm]
\hline\\[-3mm]
\includegraphics[scale=0.6]{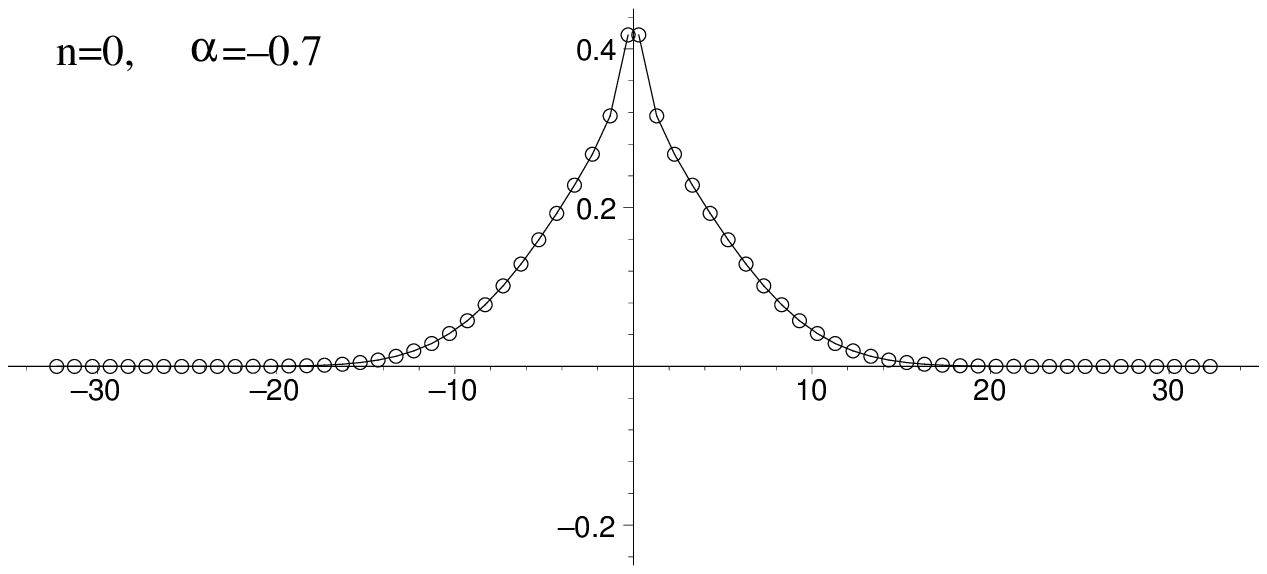} & \includegraphics[scale=0.6]{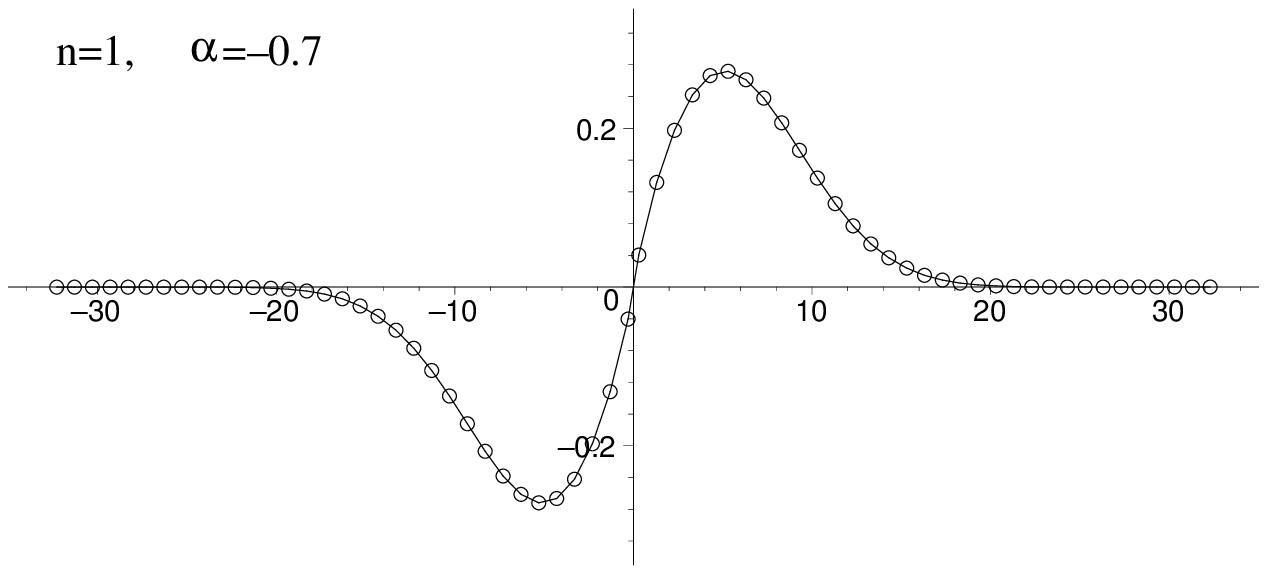} \\
\includegraphics[scale=0.6]{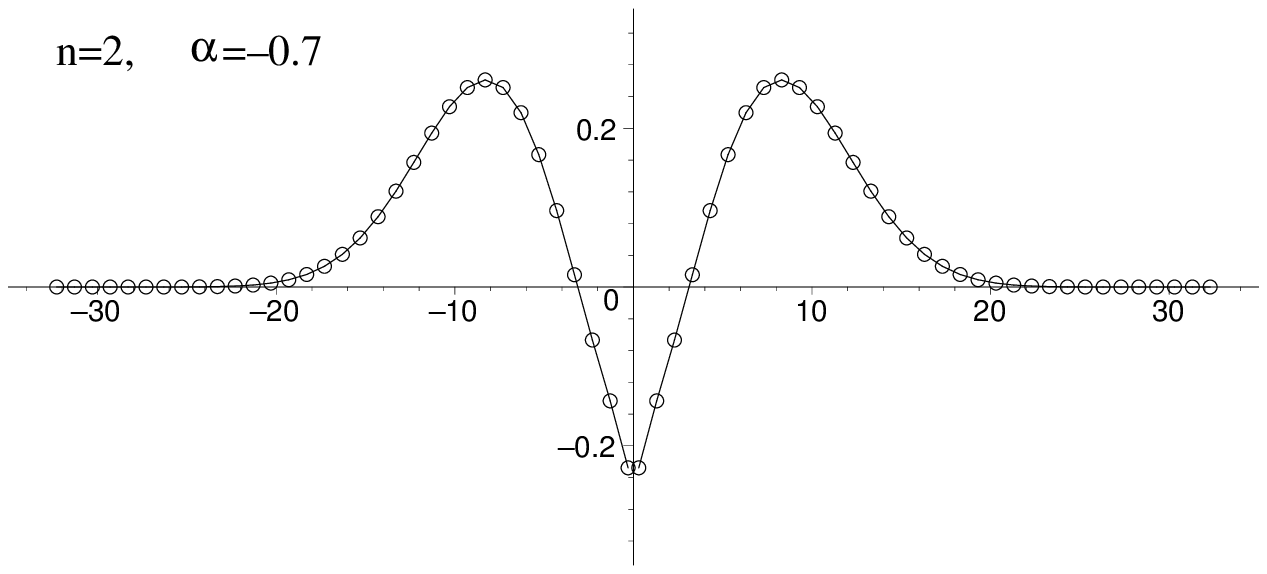} & \includegraphics[scale=0.6]{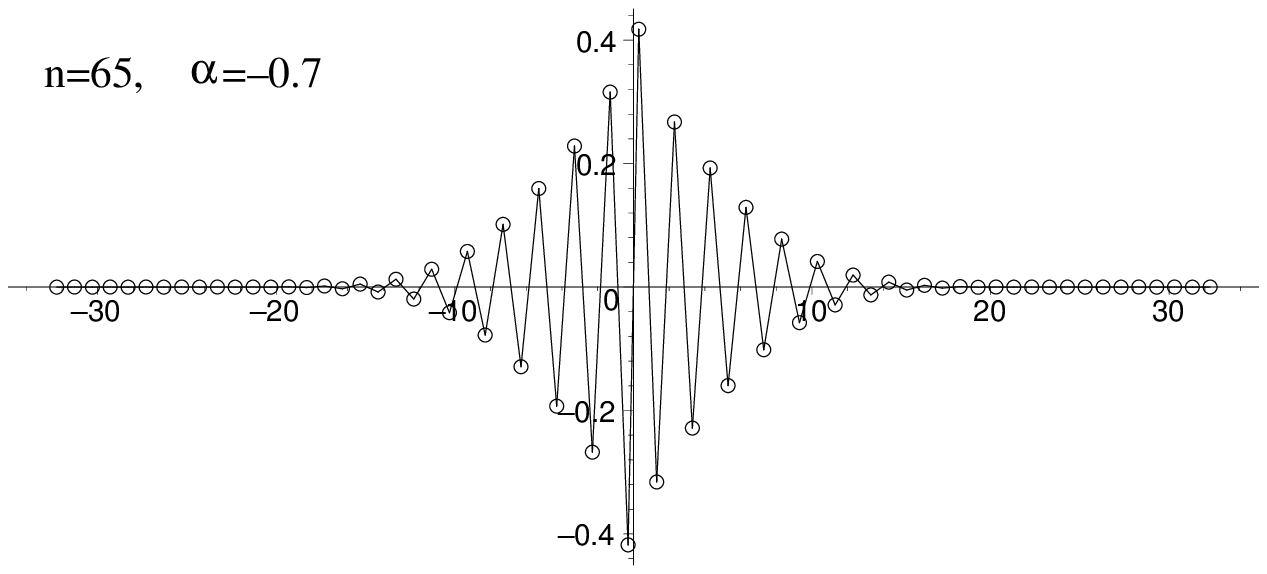} \\[-1mm]
\hline\\[-3mm]
\includegraphics[scale=0.6]{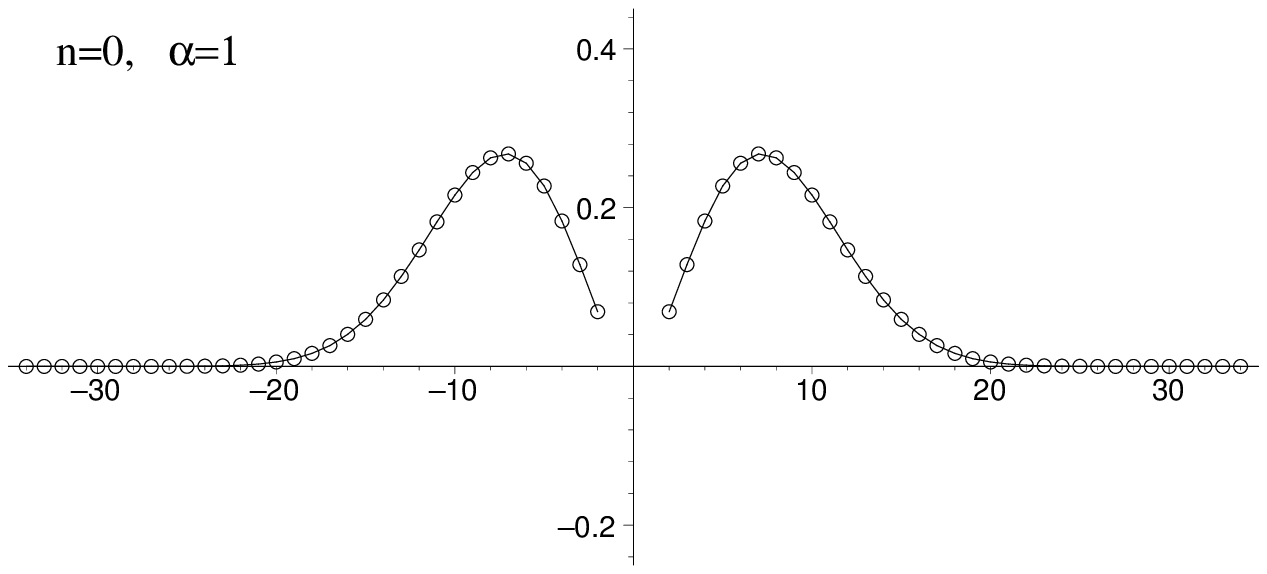} & \includegraphics[scale=0.6]{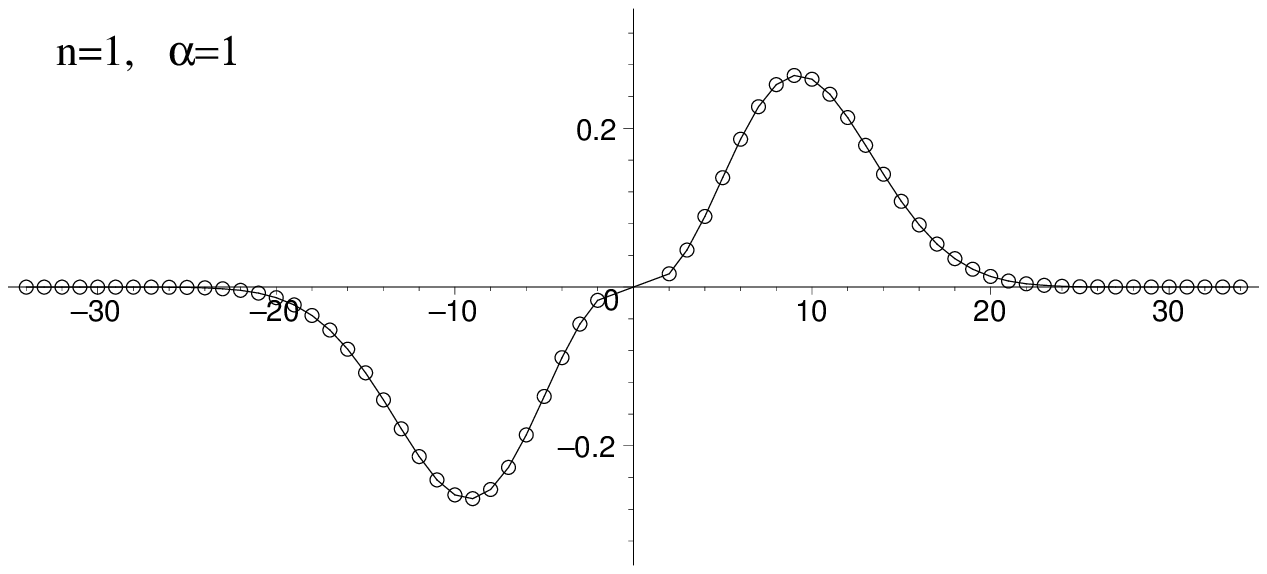} \\
\includegraphics[scale=0.6]{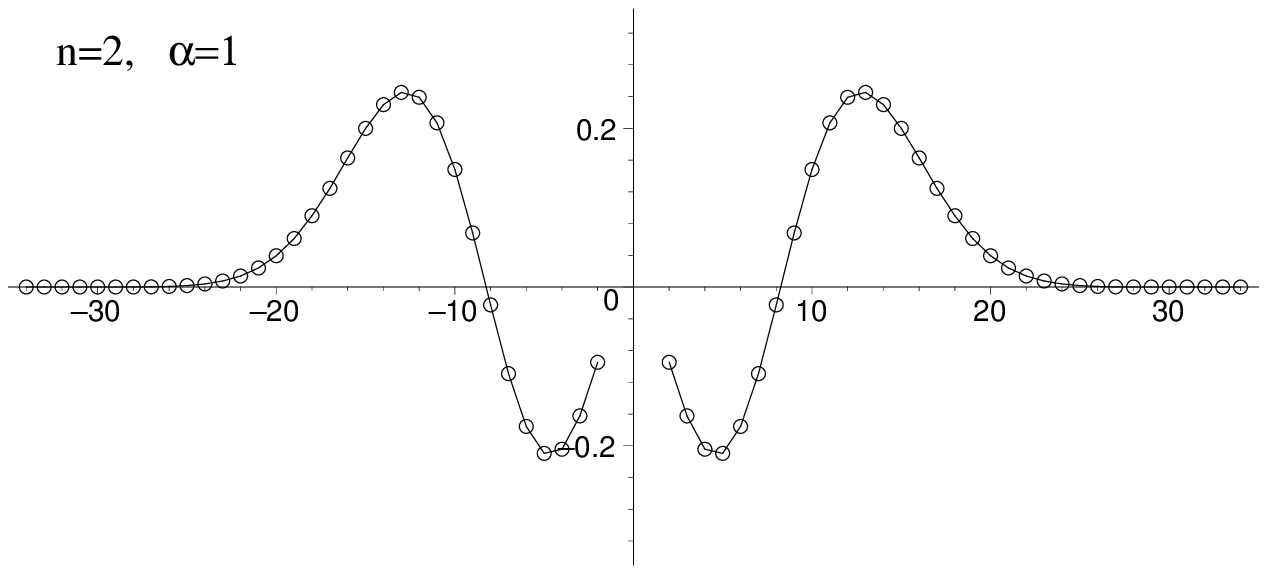} & \includegraphics[scale=0.6]{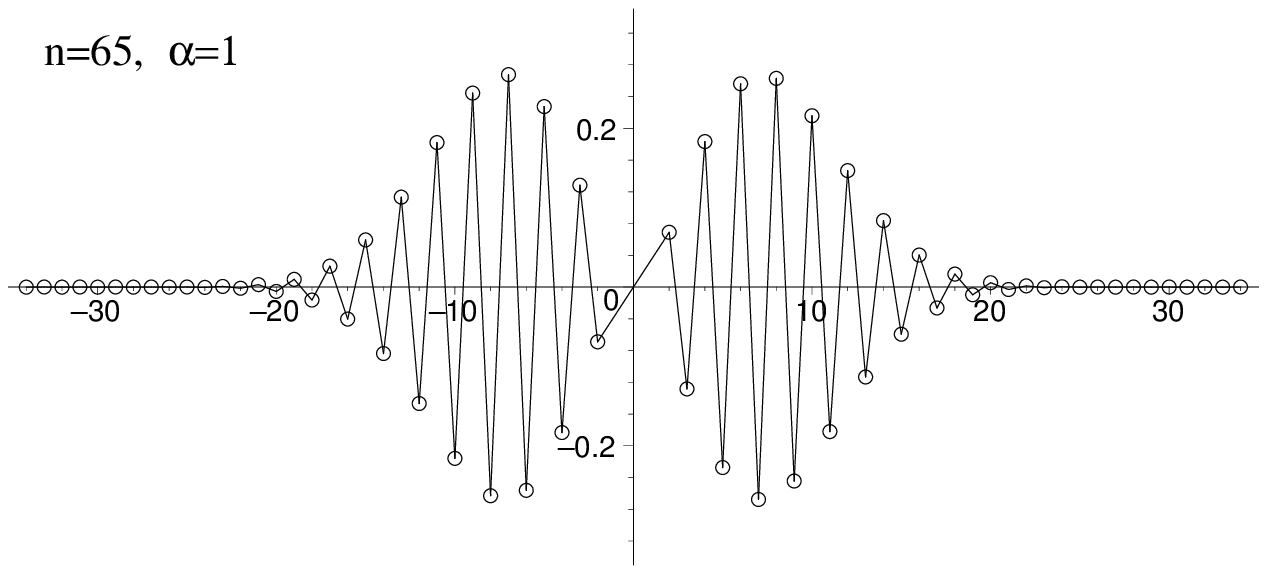} \\[-1mm]
\hline
\end{tabular} 
\caption{Plots of the discrete wavefunctions $\Phi^{(\alpha)}_n(q)$ in the representation with $j=65/2$. The four top figures are for $\alpha=-1/2$,
the middle figures for $\alpha=-0.7$, and the four bottom figures for $\alpha=1$. In each case, we plot
the wavefunctions for $n=0,1,2,65$.}
\label{fig1}
\end{figure}

It is interesting to study some plots of these discrete wavefunctions, for some values of $\alpha$.
Let us choose a fixed value of $j$, say $j=\frac{65}{2}$, and plot some of the 
wavefunctions $\Phi^{(\alpha)}_n(q)$ for various values of $\alpha$.
Since $\alpha=-\frac12$ is a special case (where $\u(2)_\alpha$ reduces to $\u(2)$), there are
three cases to be considered: $-1<\alpha<-\frac12$, $\alpha=-\frac12$ and $\alpha>-\frac12$.
In Figure~\ref{fig1} we take $\alpha=-\frac12$, $\alpha=-0.7$ and $\alpha=1$ respectively.
We also plot in each case the ground state $\Phi^{(\alpha)}_0(q)$, some low energy states $\Phi^{(\alpha)}_1(q)$ and $\Phi^{(\alpha)}_2(q)$,
and the highest energy state.

For $\alpha=-\frac12$, these plots are familiar. 
In that case, the current finite oscillator model coincides with the model of Atakishiyev {\em et al}~\cite{Atak2001,Atak2005}, based 
upon the $\su(2)$ algebra. 
It is known that in such a case, the wavefunctions $\Phi^{(-1/2)}_n(q)$ are in fact Krawtchouk functions.
This is indeed a special case of our wavefunctions, expressed as dual Hahn functions.
This follows from the fact that when $\alpha=-\frac12$ the dual Hahn polynomials, which are ${}_3F_2$ series
appearing in~\eqref{Phi-even}-\eqref{Phi-odd}, reduce to ${}_2F_1$ series
according to
\begin{align}
& {\;}_3F_2 \left( \myatop{-q+1/2,q+1/2,-n}{1/2,-j+1/2} ; 1 \right)= (-1)^n \frac{\binom{2j}{2n}}{\binom{j-1/2}{n}} 
{\;}_2F_1 \left( \myatop{-2n,-j-q}{-2j} ; 2 \right),\\
& {\;}_3F_2 \left( \myatop{-q+1/2,q+1/2,-n}{3/2,-j+1/2} ; 1 \right)= -\frac{(-1)^n}{2q} \frac{\binom{2j}{2n+1}}{\binom{j-1/2}{n}} 
{\;}_2F_1 \left( \myatop{-2n-1,-j-q}{-2j} ; 2 \right). 
\end{align}
These reductions have been given in~\cite{SV2011} and can be obtained, e.g., from~\cite[(48)]{Atak2005}. 
The ${}_2F_1$ series in the right hand side correspond to symmetric Krawtchouk polynomials
(i.e.\ Krawtchouk polynomials with $p=1/2$~\cite{Koekoek}).

Another reason why the wavefunctions $\Phi^{(-1/2)}_n(q)$ can be considered as finite oscillator wavefunctions is 
because in the limit $j\rightarrow \infty$ they yield the ordinary oscillator wavefunctions. 
To present this limit, one should also pass from a discrete position variable $q$ to a continuous position variable $x$.
This can be done by putting $q=j^{1/2} x$, and then taking the limit. 
As a consequence of this factor $j^{1/2}$, due to the discrete orthogonality relation for the functions $\Phi^{(-1/2)}_n(q)$, which should
become a continuous orthogonality relation in the limit case, one should consider the limit of $j^{1/4} \Phi^{(-1/2)}_n(q)$.
Then one obtains:
\begin{equation}
\lim_{j\rightarrow \infty} j^{1/4} \Phi^{(-1/2)}_n(j^{1/2} x ) = \frac{1}{2^{n/2}\sqrt{n!}\pi^{1/4}} H_n(x) e^{-x^2/2},
\label{Hermite}
\end{equation}
where $H_n(x)$ are the common Hermite polynomials~\cite{Koekoek,Temme}.
For the current case with $\alpha=-\frac12$, the discrete wavefunctions are Krawtchouk functions and this limit has been computed in~\cite{Atak2005}.
Expression~\eqref{Hermite} can also be deduced from our limit for general values of $\alpha$.

Now the main question is: what is the $j\rightarrow \infty$ limit of $\Phi^{(\alpha)}_n(q)$ for general $\alpha$ ($\alpha>-1$)?
To compute this, let us consider the case $n$ even and $n$ odd separately. 
For even values, one should consider expression~\eqref{Phi-even}, and compute
\begin{equation}
\lim_{j\rightarrow\infty} j^{1/4} \Phi^{(\alpha)}_{2n}(j^{1/2} x ).
\end{equation}
Since $q$ is positive in~\eqref{Phi-even}, $x$ is also positive; the total wavefunction is even so it is clear how to extend it to negative $x$-values.
The limit of the ${}_3F_2$ function in~\eqref{Phi-even} is quite easy:
\begin{equation}
\lim_{j\rightarrow \infty} {\ }_3F_2 \left( \myatop{-j^{1/2}x+\alpha+1,j^{1/2}x+\alpha+1,-n}{\alpha+1,-j+\frac12} ; 1 \right) =
{\ }_1F_1 \left( \myatop{-n}{\alpha+1} ; x^2 \right) = 
\frac{n!}{(\alpha+1)_n} L_n^{(\alpha)}(x^2),
\end{equation}
where $L_n^{(\alpha)}$ is a Laguerre polynomial~\cite{Koekoek,Temme}.
So it remains to determine
\begin{equation}
\lim_{j\rightarrow \infty} \frac{(-1)^n}{\sqrt{2}} 
\sqrt{\frac{ j^{1/2} w(n;\alpha,\alpha+1,j-\frac12)}{h(j^{1/2} x-\alpha-1;\alpha,\alpha+1,j-\frac12)}} .
\label{limwd}
\end{equation}
In order to compute this, one should replace all Pochhammer symbols in the expression of $h$ by Gamma-functions,
i.e.\ $(a)_k= \Gamma(a+k)/\Gamma(a)$. 
Then it is a matter of combining appropriate factors in the quotient of~\eqref{limwd}, 
and using Stirling's approximation for the Gamma-function~\cite{Temme}.
This leads to:
\begin{equation}
\lim_{j\rightarrow \infty} 
\frac{ j^{1/2} w(n;\alpha,\alpha+1,j-\frac12)}{2 h(j^{1/2} x-\alpha-1;\alpha,\alpha+1,j-\frac12)} =
\frac{(\alpha+1)_n}{n! \Gamma(\alpha+1)} x^{2\alpha+1} e^{-x^2}.
\end{equation}
Combining all these results, and extending it trivially to negative $x$-values, one finds:
\begin{equation}
\lim_{j\rightarrow\infty} j^{1/4} \Phi^{(\alpha)}_{2n}(j^{1/2} x ) = 
(-1)^n \sqrt{\frac{n!}{\Gamma(\alpha+n+1)}}\; |x|^{\alpha+1/2} e^{-x^2/2} L_n^{(\alpha)}(x^2).
\label{psi-even}
\end{equation}

For odd values of $n$, the limit of the ${}_3F_2$ function in~\eqref{Phi-odd} yields:
\begin{equation}
\lim_{j\rightarrow \infty} {\ }_3F_2 \left( \myatop{-j^{1/2}x+\alpha+1,j^{1/2}x+\alpha+1,-n}{\alpha+2,-j+\frac12} ; 1 \right) =
{\ }_1F_1 \left( \myatop{-n}{\alpha+2} ; x^2 \right) = 
\frac{n!}{(\alpha+2)_n} L_n^{(\alpha+1)}(x^2).
\end{equation}
Performing a similar computation as before (and extending it to negative $x$-values as an odd function of $x$), the final result is:
\begin{equation}
\lim_{j\rightarrow\infty} j^{1/4} \Phi^{(\alpha)}_{2n+1}(j^{1/2} x ) = 
(-1)^n \sqrt{\frac{n!}{\Gamma(\alpha+n+2)}}\; x |x|^{\alpha+1/2} e^{-x^2/2} L_n^{(\alpha+1)}(x^2).
\label{psi-odd}
\end{equation}
Note that for $\alpha=-1/2$, one finds indeed~\eqref{Hermite}. 

The functions in~\eqref{psi-even} and \eqref{psi-odd} are familiar: they are in fact the wavefunctions $\Psi^{(\alpha+1)}_n(x)$ of 
the parabose oscillator with parameter $a=\alpha+1>0$ (see the appendix). So we have:
\begin{equation}
\lim_{j\rightarrow\infty} j^{1/4} \Phi^{(\alpha)}_{n}(j^{1/2} x ) = \Psi^{(\alpha+1)}_n(x).
\end{equation}
In this sense, the current model can be interpreted as a finite one-dimensional parabose oscillator model.
This also explains the shape of the discrete wavefunctions plotted in Figure~1.
For $-1<\alpha<-\frac12$, the shape typically reproduces the continuous wavefunctions of the parabose
oscillator with $0<a=\alpha+1<\frac12$: see the plots for $\alpha=-0.7$ in Figure~1 and
those for $a=0.3$ in Figure~2.
For $\alpha>-\frac12$, the shape of the wavefunctions is different than for $\alpha<-\frac12$.
Now the shape is similar to those of the parabose oscillator with $a>\frac12$: compare the plots
for $\alpha=1$ in Figure~1 with those for $a=2$ in Figure~2.

\section{Discussion}

The quantum harmonic oscillator in the canonical non-relativistic case has a very simple and well known solution,
with equidistant energy spectrum and stationary states described in terms of Hermite polynomials.
It is used as a model in many applications.
However, it also has some restrictions. 
In particular, it has an infinite spectrum (which is sometimes not realistic), and its wavefunctions satisfy a continuous
orthogonality relation with infinite support.
Therefore, it is not directly applicable to describe models where only a finite number of eigenmodes can exist,
such as in optical image processing.

In this context~\cite{Atak2005} one realized that the quantum oscillator equations of motion are consistent
with commutators other than $[ \hat q, \hat p]=i$. 
Fractional Fourier transforms~\cite[Ch.~10]{Ozaktas} to signal analysis on a finite number of discrete sensors or data points
led to physical models realizing a one-dimensional finite oscillator~\cite{Atak1994,Atak1997,Atak1999b}.

Finite oscillator models are specific examples of quantum systems with a finite Hilbert space.
Such finite quantum systems have been studied from a general point of view, see the review paper~\cite{Vourdas} and references therein.
A wide variety of applications of finite quantum systems is known, of which quantum optics and quantum computing are the most popular.
Since our contribution is devoted to a model of the one-dimensional finite oscillator, primarily applications in quantum optics are worth considering.
Having this in mind, it is necessary to turn to a two-dimensional finite oscillator model and study wavefunctions on a finite (square) 
Cartesian grid~\cite{Atak2001}.
In the simplest picture, this follows just from a direct product of two one-dimensional oscillators, and the 
wavefunctions in an $xy$-plane could be described, in an obvious notation as in~\cite{Atak2001}, as 
\[
\Phi^{(\alpha_1,\alpha_2)}_{n_x,n_y}(q_x,q_y)= \Phi^{(\alpha_1)}_{n_x}(q_x) \Phi^{(\alpha_2)}_{n_y}(q_y).
\]
These Cartesian mode wavefunctions could be depicted as in~\cite[Figure~3]{Atak2001}, with however two main differences.
First of all, the ``density distribution'' depends on the $\alpha$-parameters, as in the one-dimensional plots of Figure~1.
Secondly, the ``sensor points'' of the grid are not uniformly distributed, but according to the position
spectrum given in Proposition~3.
We hope that these extra facilities open the way to more sophisticated techniques for two-dimensional signals
on square screens, in the spirit of~\cite{Wolf2010}.
This analysis, however, falls outside the scope of the present paper.

The most interesting (mathematical) aspects of finite oscillators were given in~\cite{Atak2001,Atak2001b,Atak2005}.
The model of the finite oscillator considered there is based on the Lie algebra $\su(2)$ and its representations labelled by 
an integer or half-integer~$j$.
The authors have shown that the wavefunctions of the $\su(2)$ finite oscillator are given by Krawtchouk polynomials,
and that in the limit $j\rightarrow\infty$ these discrete wavefunctions tend to the canonical oscillator
wavefunctions in terms of Hermite polynomials. 
The position operator has a finite equidistant spectrum.
These properties are summarized in the bottom part of Table~1. 

\begin{table}[htb]
\begin{center}
\begin{tabular}{ccc}
\begin{tabular}{|c|}
\hline
{\em finite $\u(2)_\alpha$ oscillator}\\
{\em (Hahn oscillator)} \\
algebra $\u(2)_\alpha$\\
discrete position spectrum with a gap\\
dual Hahn polynomial\\
\hline
\end{tabular}
 & \qquad $\myatop{j\rightarrow \infty}{\longrightarrow}$ \qquad &
\begin{tabular}{|c|}
\hline
 {\em parabose oscillator} \\
 {\em (Wigner quantum oscillator)}\\
 Lie superalgebra $\osp(1|2)$ \\
 position spectrum: $\R$ or $\R\setminus\{0\}$ \\
 Laguerre polynomial \\
 \hline
\end{tabular} \\
 & & \\ 
$\downarrow \ (\alpha=-\frac12)$ &    & $\downarrow$  ($\alpha=-\frac12$ or $a=\frac12$) \\
 & & \\
\begin{tabular}{|c|}
\hline
{\em finite $\su(2)$ oscillator}\\
{\em (Krawtchouk oscillator)}\\
Lie algebra $\su(2)$\\
discrete equidistant position spectrum \\
Krawtchouk polynomial\\
\hline
\end{tabular}
 & \qquad $\myatop{j\rightarrow \infty}{\longrightarrow}$ \qquad & 
\begin{tabular}{|c|} 
\hline
{\em canonical oscillator}\\
 oscillator algebra \\
 position spectrum: $\R$ \\
 Hermite polynomial \\
 \hline
\end{tabular} 
\end{tabular}
\end{center}
\caption{Summary of some properties of the four oscillator models appearing in this paper. 
For each of them, we give the dynamical algebra, the spectrum of the position operator, and the polynomials
appearing in the wavefunctions. The arrows indicate how to go from one model to another.}
\end{table}

In the present paper we have been able to extend the $\su(2)$ finite oscillator model by introducing an
extra parameter $\alpha>-1$. The dynamical algebra, $\u(2)_\alpha$ is a deformation of the Lie algebra $\u(2)$
extended by a parity operator. The representations of $\u(2)_\alpha$ are those of $\u(2)$ characterized by
a half-integer~$j$, but deformed by the parameter~$\alpha$.
The spectrum of the position operator has been determined: it is again finite and equidistant in steps of one unit, except that
there is a gap of size $2\alpha+2$ in the middle of the spectrum.
The wavefunctions of the $\u(2)_\alpha$ finite oscillator have been determined, and turn out to be
dual Hahn polynomials involving this parameter $\alpha$.
When $\alpha=-\frac12$, the $\su(2)$ finite oscillator model is recovered, depicted in the left part of Table~1.

We have also investigated the $j\rightarrow\infty$ limit of the $\u(2)_\alpha$ finite oscillator model. 
It is quite remarkable that under this limit, the discrete wavefunctions in terms of dual Hahn polynomials
tend to the continuous wavefunctions of the parabose oscillator~\cite{Mukunda,JSV2008} (in terms of Laguerre polynomials, sometimes
referred to as generalized Hermite polynomials). This correspondence is given in the top part of Table~1.
Note that this parabose oscillator, often referred to as the one-dimensional Wigner quantum oscillator, 
is itself an extension of the canonical oscillator. 
The parabose oscillator is described~\cite{Ganchev} in terms of the Lie superalgebra $\osp(1|2)$, with (infinite-dimensional)
representations labelled by a positive number $a$ (in our correspondence $a=\alpha+1$).
In the representation with $a=\frac12$, the parabose oscillator coincides with the canonical oscillator.

Our introduction and study of the Hahn oscillator has not only extended the by now well known
$\su(2)$ finite oscillator model, it has also unified two approaches that extend the canonical oscillator.
In this unification, given schematically in Table~1, both the parabose oscillator model and the
$\su(2)$ finite oscillator model appear as a limiting case or a special case.

Our approach is based on a deformation of the Lie algebra $\u(2)$ and its representations.
This deformation is from a very different nature than the common $q$-deformation $su_q(2)$ in the context of quantum algebras:
in that case, the common dynamical algebras become quantum algebras, and the wavefunctions 
are deformed into the corresponding $q$-functions. 
Note that $su_q(2)$ has also been considered to build a model for a finite oscillator~\cite{Ballesteros,AKW}.
In that case, the position operator has a discrete anharmonic spectrum, and the wavefunctions
are given in terms of dual $q$-Krawtchouk polynomials~\cite{AKW}.

Note, finally, the relation with a recent paper on the shifted harmonic approximation~\cite{Rowe2010}.
In that paper, the authors define in a sense discrete wavefunctions as Clebsch-Gordan coefficients
of $SU(2)$ or $SU(1,1)$. They study some plots of these discrete functions, and also show that in the 
limit these functions approach the harmonic oscillator wavefunctions.
Since $SU(2)$ or $SU(1,1)$ Clebsch-Gordan coefficients can also be expressed as ${}_3F_2(1)$
series, i.e.\ as Hahn or dual Hahn polynomials~\cite{Koornwinder,Vanderjeugt2003}, the link with the current paper is clear,
giving in a way a more fundamental reason why Clebsch-Gordan coefficients can be 
considered as discrete wavefunctions.

We hope to investigate in a more systematical manner how the algebra $\u(2)$ can be deformed,
as in~\eqref{J+J-}, in a way that is still consistent with operator actions in representations.
Such a general approach might lead to a unification at a higher level, in terms of polynomials
at a higher level in the Askey-scheme.

\section*{Appendix}

Let us summarize some aspects of the parabose oscillator, 
or the one-dimensional Wigner quantum oscillator (see also~\cite{JSV2008} for an overview).
Consider the Hamiltonian for a one-dimensional harmonic oscillator:
\begin{equation}
\label{H}
\hat H = \frac{\hat p^2}{2} + \frac{\hat q^2}{2},
\end{equation}
where $\hat p$ and $\hat q$ denote respectively the momentum and position operator 
of the system.  Wigner~\cite{Wigner} already noted that there are other 
solutions besides the canonical one if one only requires the compatibility
between the Hamilton and the Heisenberg equations (dropping the canonical
commutation relation $[\hat q,\hat p] = i$).
These compatibility conditions are:
\begin{equation}
\label{CCs}
[\hat H, \hat p] = i\hat q,\quad
[\hat H, \hat q] = -i\hat p.
\end{equation}
So, one has to find operators $\hat p$ and $\hat q$, acting in some Hilbert space,
such that the compatibility conditions~\eqref{CCs} hold, with $\hat H$ given by~\eqref{H}.
Also, in this Hilbert space $\hat p$ and $\hat q$ have to be self-adjoint.

The solutions to~\eqref{CCs} can be found by introducing two new
operators $\hat b^+$ and $\hat b^-$ (the parabose creation and annihilation operators):
\begin{equation}\label{bpm}
\hat b^\pm = \frac{1}{\sqrt{2}}(\hat q \mp i\hat p), 
\end{equation}
or equivalently
\begin{equation*}
\hat q = \frac{1}{\sqrt{2}}(\hat b^+ + \hat b^-),\quad
\hat p = \frac{i}{\sqrt{2}}(\hat b^+ - \hat b^-).
\end{equation*}
It is then easily checked that
\begin{equation}\label{Hb}
\hat H = \frac12\{\hat b^-,\hat b^+\},
\end{equation}
and that the compatibility conditions~\eqref{CCs} are equivalent with
\begin{equation}
	[\{\hat b^-,\hat b^+\}, \hat b^\pm] = \pm 2 \hat b^\pm.
	\label{CCs2}
\end{equation}

The relations~\eqref{CCs2} are in fact the defining relations of 
one pair of parabose operators $\hat b^\pm$~\cite{Green53}.
Moreover, from the self-adjointness of the position and momentum operators it 
follows that
\begin{equation}
(\hat b^\pm)^\dagger = \hat b^\mp.
	\label{bpm-adjoint}
\end{equation}
It is known that the Lie superalgebra generated by two odd elements 
$\hat b^\pm$ subject to the restriction~\eqref{CCs2} is the Lie superalgebra
$\osp(1|2)$~\cite{Ganchev}.  
So, the solutions to the problem are given by the star representations
of the Lie superalgebra $\osp(1|2)$.  These are known, and are characterized
by a positive real number $a$ and a vacuum vector $|0\rangle$, such that
\begin{equation*}
\hat b^-|0\rangle = 0,\quad
\{\hat b^-,\hat b^+\} |0\rangle = 2a |0\rangle.
\end{equation*}
The representation space can then be shown to be the Hilbert space $\ell^2(\Z_+)$
with orthonormal basis vectors $|n\rangle$ ($n\in \Z_+$) and with the following 
actions:
\begin{equation}
\begin{aligned}
	\hat b^+ |2n\rangle & = \sqrt{2(n+a)}\,|2n+1\rangle, & & \quad &
	\hat b^- |2n\rangle & = \sqrt{2n}\,|2n-1\rangle, \\ 
	\hat b^+ |2n+1\rangle & = \sqrt{2(n+1)}\,|2n+2\rangle, & &  \quad & 
	\hat b^- |2n+1\rangle & = \sqrt{2(n+a)}\,|2n\rangle,  
\end{aligned}
	\label{bpm-actions}
\end{equation}
from which it immediately follows that
\begin{equation}
\{ \hat b^-, \hat b^+ \}|n\rangle = 2(n+a)\,|n\rangle.
	\label{action-anticomm}
\end{equation}
The energy spectrum follows from~\eqref{action-anticomm} and~\eqref{Hb}:
\begin{equation*}
\hat H |n\rangle = (n+a)\,|n\rangle,
\end{equation*}
so one has an equidistant energy spectrum with ground level given by $a$. 
This, and the explicit action of the commutator action $[\hat p, \hat q]$ on basis vectors $|n\rangle$, 
confirms that only $a=1/2$ yields the canonical solution.

To construct the wave functions, one can work as follows.
Using $\hat q=(\hat b^+ + \hat b^-)/\sqrt{2}$, one finds now from~(\ref{bpm-actions})
\begin{align}
& \hat q | 2n\rangle = \frac{1}{\sqrt{2}}( \sqrt{2n} | 2n-1\rangle + \sqrt{2(n+a)} | 2n+1\rangle ),\nn \\
& \hat q | 2n+1\rangle = \frac{1}{\sqrt{2}}( \sqrt{2(n+a)} | 2n\rangle + \sqrt{2(n+1)} | 2n+2\rangle ).
\label{actionra}
\end{align}
This means that $\hat q$ is (or extends to) an unbounded Jacobi operator
on $\ell^2(\Z_+)$. This Jacobi operator corresponds to the generalized Hermite
polynomials, and the spectrum of $\hat q$ is $\R$~\cite{Regniers2010}. 
Consider the formal eigenvectors $v(x)$ 
of $\hat q$, for the eigenvalue $x$, and write these again as
\begin{equation}
v(x)=\sum_{n=0}^\infty \Psi^{(a)}_n(x) | n\rangle.
\end{equation}
The equation $\hat q \, v(x) = x\, v(x)$ leads, using (\ref{actionra}), to a set of
recurrence relations for the coefficients $\Psi^{(a)}_n(x)$.
Solving these recurrence relations explicitly, taking into account the
normalization condition, one finds, in terms of Laguerre polynomials:
\begin{align}
	\Psi_{2n}^{(a)}(x) & = (-1)^n  \sqrt {\frac{n!}{\Gamma( n + a) } }\, 
	|x|^{a-1/2}\,  e^{-x^2/2} L_n^{(a-1)}(x^2), \nn \\
\Psi_{2n+1}^{(a)}(x) & = (-1)^n  \sqrt {\frac{n!}{\Gamma( n + a+1) } }\,  
|x|^{a-1/2}\, e^{-x^2/2} x L_n^{(a)}(x^2).
\label{wave}
\end{align}
Since these coefficients have an interpretation as the position wavefunctions
of the Wigner oscillator, we are done.
Alternatively, one can work in the position representation,
where the operator $\hat q$ is still represented by \lq\lq multiplication by $x$\rq\rq,
and the operator $\hat p$ has a realization as $-i\frac{d}{dx}$ plus an extra term 
(depending on $a$)~\cite[Chapter 23]{Ohnuki}.  Using this realization the time-independent 
Schr\"odinger equation can be solved, also yielding the expressions~\eqref{wave}~\cite{Mukunda}.

\begin{figure}[htb]
\begin{tabular}{ccc}
\hline\\
\includegraphics[scale=0.6]{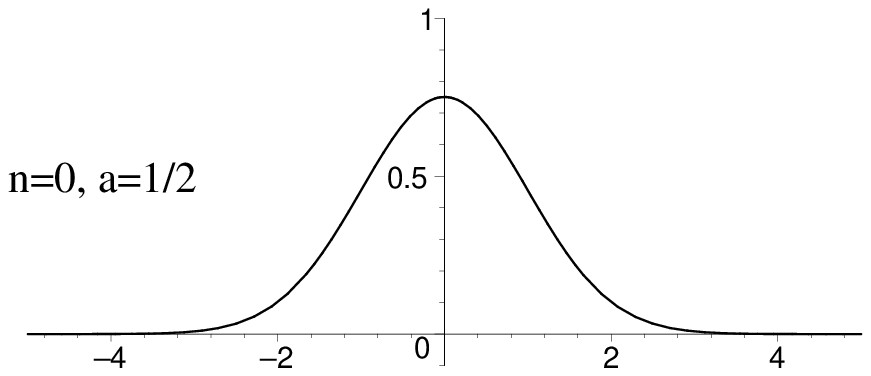} & \includegraphics[scale=0.6]{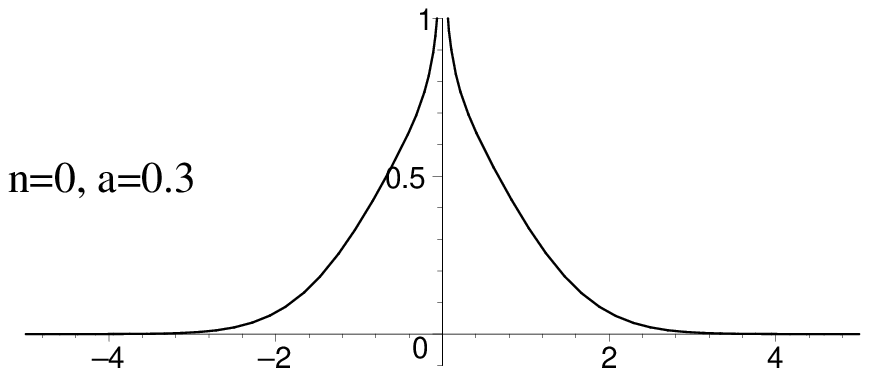} & \includegraphics[scale=0.6]{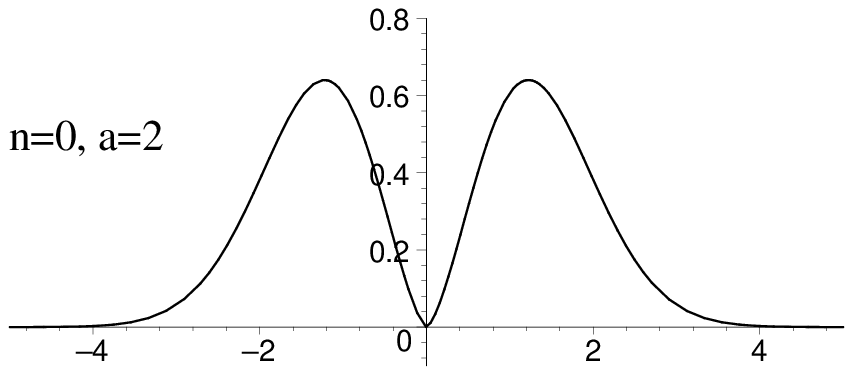} \\ \includegraphics[scale=0.6]{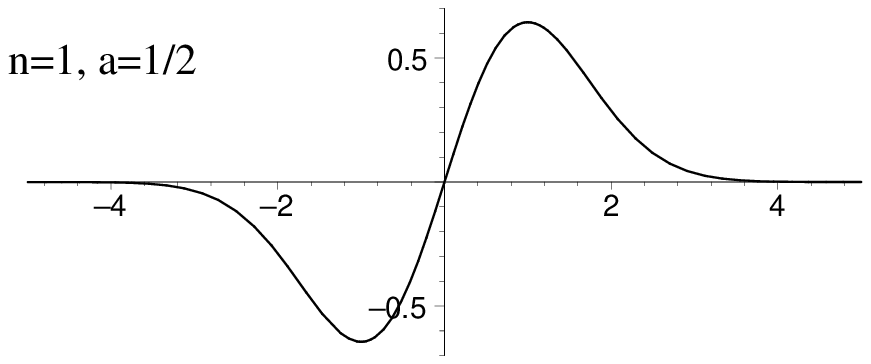} & \includegraphics[scale=0.6]{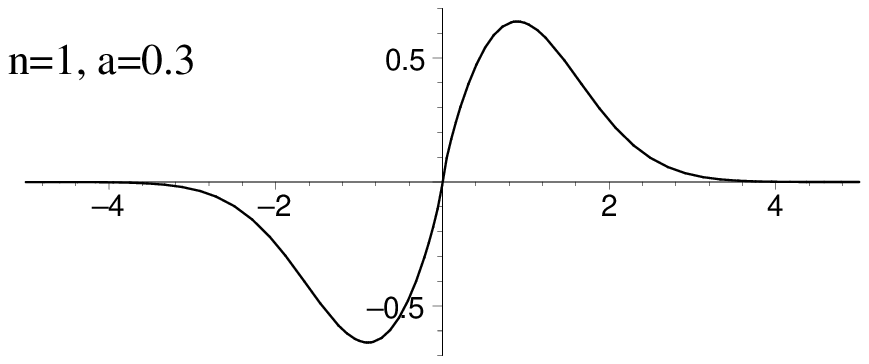} & \includegraphics[scale=0.6]{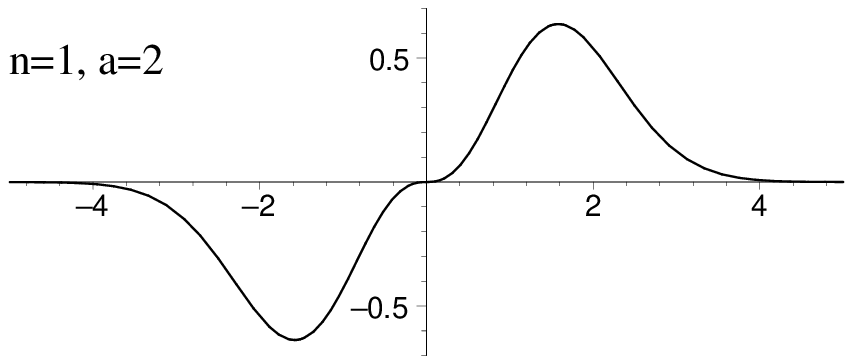} \\
\includegraphics[scale=0.6]{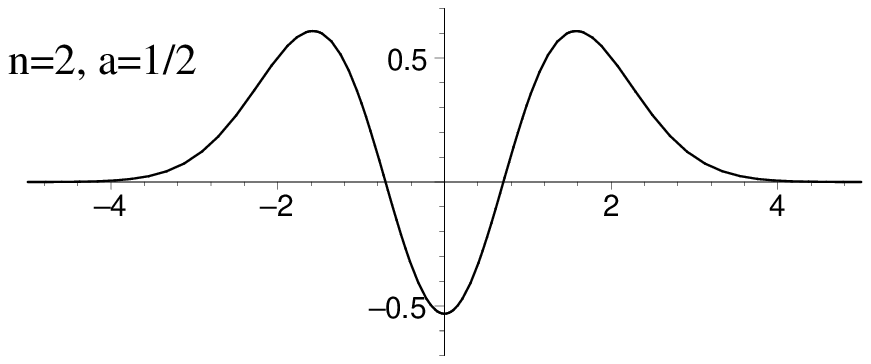} & \includegraphics[scale=0.6]{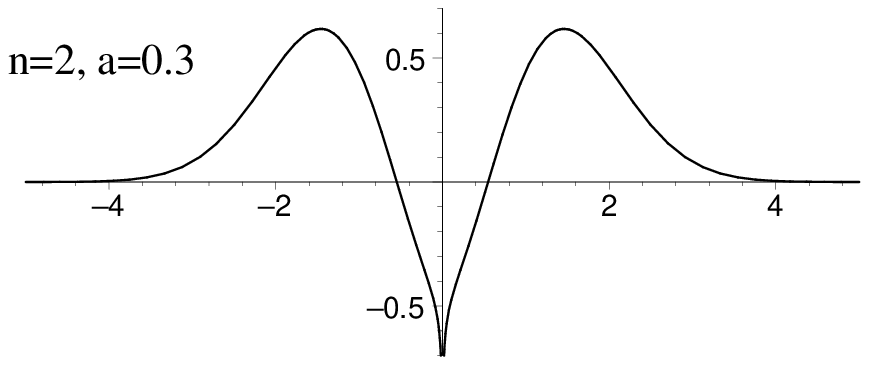} & \includegraphics[scale=0.6]{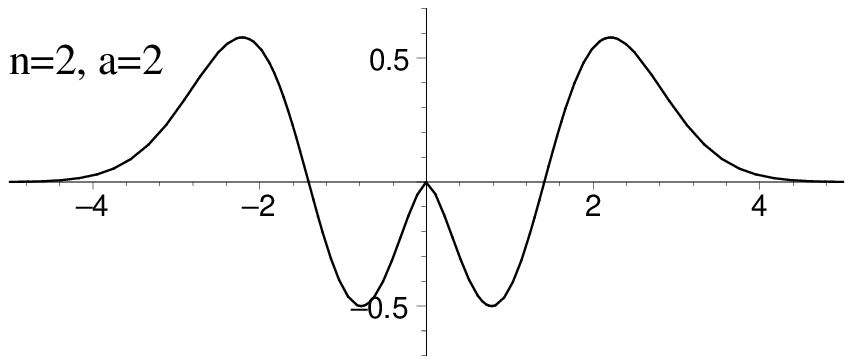}   \\
\hline
\end{tabular} 
\caption{Plots of the parabose oscillator wavefunctions $\Psi^{(a)}_n(x)$. The three figures on the left
are for $a=1/2$ and correspond to the canonical case;
the figures in the middle are for $a=0.3$, and three figures on the right are for $a=2$. In each case, we plot
the wavefunctions for $n=0,1,2$.}
\label{fig2}
\end{figure}

To compare with the discrete case considered earlier, let us also plot in Figure~\ref{fig2} the wavefunctions
$\Psi_{n}^{(a)}(x)$ for $n=0,1,2$ and for $a=0.3$, $a=\frac12$ and $a=2$ (corresponding to
$\alpha=-0.7$, $\alpha=-\frac12$ and $\alpha=1$ respectively).
Note that the shape of the discrete parabose oscillator wavefunctions are similar to
those of the continuous parabose oscillator, a fact explained by the limit relation.

\section*{Acknowledgments}
E.I.~Jafarov was supported by a postdoc fellowship from the Azerbaijan National Academy of Sciences.
N.I.~Stoilova was supported by project P6/02 of the Interuniversity Attraction Poles Programme (Belgian State -- 
Belgian Science Policy).

\end{document}